\documentclass[aps,showpacs,preprintnumbers,nofootinbibt,twocolumn]{revtex4}
\usepackage{graphicx}

\begin{document}

\title{The matter Lagrangian and the energy-momentum tensor in modified gravity with non-minimal
coupling between matter and geometry}
\author{T. Harko}
\email{harko@hkucc.hku.hk}
\affiliation{Department of Physics and
Center for Theoretical and Computational Physics, The University
of Hong Kong, Pok Fu Lam Road, Hong Kong, P. R. China}

\begin{abstract}
We show that in modified $f(R)$ type gravity models with non-minimal
coupling between matter and geometry, both the matter Lagrangian, and the
energy-momentum tensor, are completely and uniquely determined by the form
of the coupling. This result is obtained by using the variational
formulation for the derivation of the equations of motion in the modified gravity models with
geometry-matter coupling, and the Newtonian limit for a fluid obeying a
barotropic equation of state. The corresponding energy-momentum tensor of
the matter in modified gravity models with non-minimal coupling is more
general than the usual general-relativistic energy-momentum tensor for
perfect fluids, and it contains a supplementary, equation of state dependent term, which could be related to
the elastic stresses in the body, or to other forms of internal energy.
Therefore, the extra-force induced by the coupling between matter and
geometry never vanishes as a consequence of the thermodynamic properties of
the system, or for a specific choice of the matter Lagrangian, and it is non-zero in the case of a fluid of dust particles.
\end{abstract}


\date{\today}

\maketitle

\section{Introduction}

A very promising way to explain the recent observational data \cite{Ri98,
PeRa03} on the acceleration of the Universe and on dark matter is to
assume that at large scales the Einstein gravity model of general relativity
breaks down, and a more general action describes the gravitational field.
Theoretical models in which the standard Einstein-Hilbert action is replaced
by an arbitrary function of the Ricci scalar $R$, first proposed in \cite
{Bu70}, have been extensively investigated lately. Cosmic acceleration can
be explained by $f(R)$ gravity \cite{Carroll:2003wy}, and the conditions of
viable cosmological models have been derived in \cite{viablemodels}. In the
context of the Solar System regime, severe weak field constraints seem to
rule out most of the models proposed so far \cite{solartests,Olmo07},
although viable models do exist \cite
{Hu:2007nk,solartests2,Sawicki:2007tf,Amendola:2007nt}. The possibility that
the galactic dynamic of massive test particles can be understood without the
need for dark matter was also considered in the framework of $\ f(R)$
gravity models \cite{Cap2,Borowiec:2006qr,Mar1,Boehmer:2007kx,Bohmer:2007fh}%
. For a review of $f(R)$ generalized gravity models see \cite{SoFa08}.

A generalization of the $f(R)$ gravity theories was proposed in \cite
{Bertolami:2007gv} by including in the theory an explicit coupling of an
arbitrary function of the Ricci scalar $R$ with the matter Lagrangian
density $L_{m}$. As a result of the coupling the motion of the massive
particles is non-geodesic, and an extra force, orthogonal to the
four-velocity, arises. The connections with MOND and the Pioneer anomaly
were also explored. This model was extended to the case of the arbitrary
couplings in both geometry and matter in \cite{ha08}. The implications of
the non-minimal coupling on the stellar equilibrium were investigated in
\cite{Bertolami:2007vu}, where constraints on the coupling were also
obtained. An inequality which expresses a necessary and sufficient condition
to avoid the Dolgov-Kawasaki instability for the model was derived in \cite
{Fa07}. The relation between the model with geometry-matter coupling and
ordinary scalar-tensor gravity, or scalar-tensor theories which include
non-standard couplings between the scalar and matter was studied in \cite
{SoFa08a}. In the specific case where both the action and the coupling are
linear in $R$ the action leads to a theory of gravity which includes higher
order derivatives of the matter fields without introducing more dynamics in
the gravity sector \cite{So08}. The equivalence between a scalar theory and
the model with the non-minimal coupling of the scalar curvature and matter
was considered in \cite{BePa08}. This equivalence allows for the calculation
of the PPN parameters $\beta $ and $\gamma $, which may lead to a better
understanding of the weak-field limit of $f(R)$ theories. The equations of
motion of test bodies in the nonminimal coupling model by means of a
multipole method were derived in \cite{Pu08}. The energy conditions and
the stability of the model under the Dolgov-Kawasaki criterion were studied
in \cite{Be09}. For a review of modified $f(R)$ gravity with geometry-matter
coupling see \cite{rev}.

The extra force in the $f(R)$ gravity with non-minimal coupling has the
intriguing property that it depends on both the matter Lagrangian, and the
energy-momentum tensor. It was pointed out in \cite{SoFa08a} that adopting
the standard Lagrangian density ${\cal{L}}_{m}^{(1)}=P$, where $P$ is the pressure,
the extra force vanishes in the case of dust. Different forms for the matter
Lagrangian density ${\cal{L}}_{m}$, and the resulting extra-force, were considered
in \cite{BeLoPa08}, and it was shown that the more natural form for ${\cal{L}}_{m}$,
${\cal{L}}_{m}^{(2)}=-\rho $, does not imply the vanishing of the extra-force. The
impact on the classical equivalence between different Lagrangian
descriptions of a perfect fluid was also analyzed. Recently, the problem of
the two Lagrangian descriptions of a perfect fluid, and its effect on the
extra force, was reconsidered in \cite{Fa09}. The main conclusion of this
study is that'' It is a fact that by choosing ${\cal{L}}_1=P$ there is no extra force
on a dust fluid, and it is equally undeniable that by choosing ${\cal{L}}_{2}=-\rho $
there will be such a force, which may ultimately provide an alternative to
dark matter.'' It is also pointed out that ''as soon as this fluid is
coupled explicitly to gravity ...the two Lagrangian densities cease to be
equivalent'' \cite{Fa09}.

It is the purpose of the present paper to show that in the $f(R)$ gravity
with non-minimal coupling the matter Lagrangian, and the corresponding
energy-momentum tensor, are not model-independent quantities, but they are
completely and uniquely determined by the nature of the coupling between
matter and geometry. This result can be obtained by deriving first the
equations of motion in the modified gravity model from a variational principle, and then considering the
Newtonian limit of the particle action for a fluid obeying a barotropic equation of
state. The corresponding energy-momentum tensor of the matter is more
general than the usual general-relativistic energy-momentum tensor for perfect fluids, and it
contains a supplementary term that may be related to the elastic stresses
in the body, or to other sources of internal energy. The matter Lagrangian can be expressed either in terms of the
density, or in terms of the pressure, and in both representations the
physical description of the system is equivalent. Therefore the presence (or
absence) of the extra-force is independent of the specific form of the
matter Lagrangian, and it never vanishes, except the case of (un)physical
systems with zero sound speed. In particular, in the case of dust particles, the extra-force is always non-zero.

The present paper is organized as follows. The matter Lagrangian and the
energy-momentum tensor in $f(R)$ gravity with non-minimal coupling are
derived in Section II. We discuss and conclude our results in Section III. In the present paper we use the Landau-Lifshitz \cite{LaLi} sign conventions and definitions, and the natural system of units with $c=8\pi G=1$.

\section{Matter and energy-momentum tensor in modified gravity with
linear coupling between geometry and matter}

The action for the modified theories of gravity proposed in  \cite
{Bertolami:2007gv} takes the following form
\begin{equation}
S=\int \left\{ \frac{1}{2}f_{1}(R)+\left[ 1+\lambda f_{2}(R)\right]
L_{m}\right\} \sqrt{-g}\;d^{4}x~,
\end{equation}
where $f_{i}(R)$ (with $i=1,2$) are arbitrary functions of the Ricci scalar $%
R$, and $L_{m}$ is the Lagrangian density corresponding to matter. The
strength of the interaction between $f_{2}(R)$ and the matter Lagrangian is
characterized by a coupling constant $\lambda $. We define the
energy-momentum tensor of the matter as \cite{LaLi}
\begin{equation}
T_{\mu \nu }=-\frac{2}{\sqrt{-g}}\left[ \frac{\partial \left( \sqrt{-g}%
L_{m}\right) }{\partial g^{\mu \nu }}-\frac{\partial }{\partial x^{\lambda }}%
\frac{\partial \left( \sqrt{-g}L_{m}\right) }{\partial \left( \partial
g^{\mu \nu }/\partial x^{\lambda }\right) }\right] .
\end{equation}

By assuming that the Lagrangian density $L_{m}$ of the matter depends only
on the metric tensor components $g_{\mu \nu }$, and not on its derivatives,
we obtain $T_{\mu \nu }=L_{m}g_{\mu \nu }-2\partial L_{m}/\partial g^{\mu
\nu }$. By taking into account the explicit form of the field equations one
obtains for the covariant divergence of the energy-momentum tensor the
equation \cite{Bertolami:2007gv}
\begin{equation}
\nabla ^{\mu }T_{\mu \nu }=2\left\{\nabla ^{\mu }\ln \left[ 1+\lambda f_{2}(R)%
\right] \right\}\frac{\partial L_{m}}{\partial g^{\mu \nu }}.  \label{cons1}
\end{equation}

As a specific example of generalized gravity models with linear
matter-geometry coupling, we consider the case in which the matter, assumed
to be a perfect thermodynamic fluid, obeys a barotropic equation of state,
with the thermodynamic pressure $p$ being a function of the {\it rest mass density of the matter} (for short: {\it matter density})  $\rho $
only, so that $p=p\left( \rho \right) $. In this case, the matter Lagrangian
density, which in the general case could be a function of both density and
pressure, $L_{m}=L_{m}\left( \rho ,p\right) $, or of only one of the
thermodynamic parameters, becomes an arbitrary function of the density of
the matter $\rho $ only, so that $L_{m}=L_{m}\left( \rho \right) $. Then the
energy-momentum tensor of the matter is given by
\begin{equation}
T^{\mu \nu }=\rho \frac{dL_{m}}{d\rho }u^{\mu }u^{\nu }+\left( L_{m}-\rho
\frac{dL_{m}}{d\rho }\right) g^{\mu \nu },  \label{tens}
\end{equation}
where the four-velocity $u^{\mu }=dx^{\mu }/ds$ satisfies the condition $%
g^{\mu \nu }u_{\mu }u_{\nu }=1$. To obtain Eq.~(\ref{tens}) we have imposed
the condition of the conservation of the matter current, $\nabla _{\nu
}\left( \rho u^{\nu }\right) =0$, and we have used the relation $\delta \rho
=\left( 1/2\right) \rho \left( g_{\mu \nu }-u_{\mu }u_{\nu }\right) \delta
g^{\mu \nu }$, whose proof is given in the Appendix. With the use of the
identity $u^{\nu }\nabla _{\nu }u^{\mu }=d^{2}x^{\mu }/ds^{2}+\Gamma _{\nu
\lambda }^{\mu }u^{\nu }u^{\lambda }$, from Eqs.~(\ref{cons1}) and (\ref
{tens}) we obtain the equation of motion of a test fluid in the modified
gravity model with linear coupling between matter and geometry as
\begin{equation}
\frac{d^{2}x^{\mu }}{ds^{2}}+\Gamma _{\nu \lambda }^{\mu }u^{\nu }u^{\lambda
}=f^{\mu },  \label{eqmot}
\end{equation}
where
\begin{equation}
f^{\mu }=-\nabla _{\nu }\ln \left\{ \left[ 1+\lambda f_{2}(R)\right] \frac{%
dL_{m}\left( \rho \right) }{d\rho }\right\} \left( u^{\mu }u^{\nu }-g^{\mu
\nu }\right) .
\end{equation}

The extra-force $f^{\mu }$, generated due to the presence of the coupling
between matter and geometry, is perpendicular to the four-velocity, $f^{\mu
}u_{\mu }=0$. The equation of motion Eq.~(\ref{eqmot}) can be obtained from
the variational principle
\begin{equation}
\delta S_{p}=\delta \int L_{p}ds=\delta \int \sqrt{Q}\sqrt{g_{\mu \nu
}u^{\mu }u^{\nu }}ds=0,  \label{actpart}
\end{equation}
where $S_{p}$ and $L_{p}=\sqrt{Q}\sqrt{g_{\mu \nu }u^{\mu }u^{\nu }}$ are
the action and the Lagrangian density for the test particles, respectively,
and
\begin{equation}
\sqrt{Q}=\left[ 1+\lambda f_{2}(R)\right] \frac{dL_{m}\left( \rho \right) }{%
d\rho }.  \label{Q}
\end{equation}

To prove this result we start with the Lagrange equations corresponding to
the action~(\ref{actpart}),
\begin{equation}
\frac{d}{ds}\left( \frac{\partial L_{p}}{\partial u^{\lambda }}\right) -%
\frac{\partial L_{p}}{\partial x^{\lambda }}=0.
\end{equation}

Since $\partial L_{p}/\partial u^{\lambda }=\sqrt{Q}u_{\lambda }$ and $%
\partial L_{p}/\partial x^{\lambda }=\left( 1/2\right) \sqrt{Q}g_{\mu \nu
,\lambda }u^{\mu }u^{\nu }+\left( 1/2\right) Q_{,\lambda }/Q$, a
straightforward calculation gives the equations of motion of the particle as
\begin{equation}
\frac{d^{2}x^{\mu }}{ds^{2}}+\Gamma _{\nu \lambda }^{\mu }u^{\nu }u^{\lambda
}+\left( u^{\mu }u^{\nu }-g^{\mu \nu }\right) \nabla _{\nu }\ln \sqrt{Q}=0.
\end{equation}
By simple identification with the equation of motion of the modified gravity
model with linear matter-geometry coupling, given by Eq.~(\ref{eqmot}), we
obtain the explicit form of $\sqrt{Q}$, as given by Eq.~(\ref{Q}). When $\sqrt{Q}\rightarrow 1$ we reobtain the standard general relativistic equation for geodesic motion.

The variational principle~(\ref{actpart}) can be used to study the Newtonian
limit of the model. In the limit of the weak gravitational fields, $%
ds\approx \sqrt{1+2\phi -\vec{v}^{2}}dt\approx \left( 1+\phi -\vec{v}%
^{2}/2\right) dt$, where $\phi $ is the Newtonian potential and $\vec{v}$ is
the usual tridimensional velocity of the fluid. By representing the function
$\sqrt{Q}$ as
\begin{equation}
\sqrt{Q}=\frac{dL_{m}\left( \rho \right) }{d\rho }+\lambda f_{2}(R)\frac{%
dL_{m}\left( \rho \right) }{d\rho },
\end{equation}
in the first order of approximation the equations of motion of the fluid can
be obtained from the variational principle
\begin{equation}
\delta \int \left[ \frac{dL_{m}\left( \rho \right) }{d\rho }+\lambda f_{2}(R)%
\frac{dL_{m}\left( \rho \right) }{d\rho }+\phi -\frac{\vec{v}^{2}}{2}\right]
dt=0,
\end{equation}
and are given by
\begin{equation}
\vec{a}=-\nabla \phi -\nabla \frac{dL_{m}\left( \rho \right) }{d\rho }%
-\nabla U_{E}=\vec{a}_{N}+\vec{a}_{H}+\vec{a}_{E},
\end{equation}
where $\vec{a}$ is the total acceleration of the system, $\vec{a}%
_{N}=-\nabla \phi $ is the Newtonian gravitational acceleration, and $\vec{a}%
_{E}=-\nabla U_{E}=-\lambda \nabla \left[ f_{2}(R)dL_{m}\left( \rho \right)
/d\rho \right] $ is a supplementary acceleration induced due to the coupling
between matter and geometry. As for the term $\vec{a}_{H}=-\nabla \left[
dL_{m}\left( \rho \right) /d\rho \right] $, it has to be identified with the
hydrodynamic acceleration term in the perfect fluid Euler equation,
\begin{equation}
\vec{a}_{H}=-\nabla \frac{dL_{m}\left( \rho \right) }{d\rho }=-\nabla
\int_{\rho _{0}}^{\rho }\frac{dp}{d\rho }\frac{d\rho }{\rho },
\end{equation}
where $\rho _{0}$, an integration constant, plays the role of a limiting
density. Hence the matter Lagrangian can be obtained by a simple integration
as
\begin{equation}
L_{m}\left( \rho \right) =\rho \left[ 1+\Pi \left( \rho \right) \right]
-\int_{p_{0}}^{p}dp,  \label{Lm}
\end{equation}
where $\Pi \left( \rho \right) =\int_{p_{0}}^{p}dp/\rho $, and we have
normalized an arbitrary integration constant to one. $p_{0}$ is an
integration constant, or a limiting pressure. The corresponding
energy-momentum tensor of the matter is given by
\begin{equation}
T^{\mu \nu }=\left\{ \rho \left[ 1+\Phi \left( \rho \right) \right] +p\left(
\rho \right) \right\} u^{\mu }u^{\nu }-p\left( \rho \right) g^{\mu \nu },
\label{tens1}
\end{equation}
respectively, where
\begin{equation}
\Phi \left( \rho \right) =\int_{\rho _{0}}^{\rho }\frac{p}{\rho ^{2}}d\rho
=\Pi \left( \rho \right) -\frac{p\left( \rho \right) }{\rho },
\end{equation}
and with all the constant terms included in the definition of $p$. By introducing the energy density of the body according to the definition $\varepsilon=\rho \left[ 1+\Phi \left( \rho \right) \right]$, the energy-momentum tensor of a test fluid can be written in the modified gravity models with geometry-matter coupling in a form similar to the standard general relativistic case, $T^{\mu \nu }=\left[\varepsilon \left(\rho \right)+p\left(\rho \right)\right] u^{\mu }u^{\nu }-p\left( \rho \right) g^{\mu \nu }$

From a physical point of view $\Phi \left( \rho \right) $ can be interpreted
as the elastic (deformation) potential energy of the body, and therefore
Eq.~(\ref{tens1}) corresponds to the energy-momentum tensor of a
compressible elastic isotropic system. The matter Lagrangian can also be
written in the simpler form $L_{m}\left( \rho \right) =\rho \Phi \left( \rho
\right) $.

If the pressure does not have a thermodynamic or radiative component one can
take $p_{0}=0$. If the pressure is a constant background quantity,
independent of the density, so that $p=p_{0}$, then $L_{m}\left( \rho
\right) =\rho $, and the energy-momentum tensor of the matter takes the form
corresponding to dust, $T^{\mu \nu }=\rho u^{\mu }u^{\nu }$.

\section{Discussions and final remarks}

In the present paper we have shown that in the $f(R)$ gravity with
non-minimal coupling the matter Lagrangian and the energy-momentum tensor
can be obtained uniquely and consistently from the form of the coupling
between matter and geometry. The coupling completely fixes both the matter
Lagrangian, and the energy-momentum tensor. Since the matter is supposed to
obey a barotropic equation of state, this result is independent of the
concrete representation of the matter Lagrangian in terms of the
thermodynamic quantities. The same results are obtained by assuming $%
L_{m}=L_{m}\left( p\right) $ - due to the equation of state $\rho $ and $p$
are freely interchangeable thermodynamic quantities, and the Lagrangians
expressed in terms of $\rho $ and $p$ only are completely equivalent. More
general situations, in which the density and pressure are functions of the
particle number and temperature, respectively, and the equation of state is
given in a parametric form, can be analyzed in a similar way.

The form of
the matter Lagrangian, and the energy-momentum tensor, are strongly
dependent on the equation of state. For example, if the barotropic equation
of state is linear, $p=\left( \gamma -1\right) \rho $, $\gamma =$ constant, $%
1\leq \gamma \leq 2$, then
\begin{equation}
L_{m}\left( \rho \right) =\rho \left\{ 1+\left(
\gamma -1\right) \left[  \ln \left( \frac{\rho }{\rho _{0}}\right) -1 %
\right] \right\},
\end{equation}
 and $\Phi \left( \rho \right) =\left( \gamma -1\right)
\ln \left( \rho /\rho _{0}\right) $, respectively. In the case of a
polytropic equation of state $p=K\rho ^{1+1/n}$, $K,n=$constant, we obtain
\begin{equation}
L_{m}\left( \rho \right) =\rho +K\left(\frac{n^{2}}{n+1}-1\right)
\rho ^{1+1/n},
\end{equation}
and $\Phi \left( \rho \right) =Kn\rho ^{1+1/n}=np\left( \rho
\right) $, respectively, where we have taken for simplicity $\rho
_{0}=p_{0}=0$. For a fluid satisfying the ideal gas equation of state $%
p=k_{B}\rho T/\mu $, where $k_{B}$ is Boltzmann's constant, $T$ is the
temperature, and $\mu $ is the mean molecular weight, we obtain
\begin{equation}
L_{m}\left( \rho \right) =\rho \left\{ 1+\frac{k_{B}T}{\mu }\left[
\ln \left( \frac{\rho }{\rho _{0}}\right) -1\right] \right\} +p_{0}.
\end{equation}
In the case of
a physical system satisfying the ideal gas equation of state, the
extra-acceleration induced by the presence of the non-minimal coupling
between matter and geometry is given by
\begin{equation}
\vec{a}_{E}\approx -\lambda \frac{k_{B}T}{\mu }\nabla \left[ f_{2}\left( R\right) \ln
\frac{\rho }{\rho _{0}}\right] ,
\end{equation}
and it is proportional to the temperature of the fluid. It is also
interesting to note that the limiting density and pressure $\rho _{0}$ and $%
p_{0}$ generate in the energy-momentum tensor some extra constant terms,
which may be interpreted as the dark energy.

In conclusion, the extra-force induced by the coupling between matter and
geometry does not vanish for any specific choices of the matter Lagrangian.
In the case of the dust, with $p=0$, the extra force is given by
\begin{equation}
f^{\mu }=-\nabla _{\nu }\ln \left[ 1+\lambda f_{2}(R)\right] \left( u^{\mu
}u^{\nu }-g^{\mu \nu }\right) ,
\end{equation}
and it is independent on the thermodynamic properties of the system, being
completely determined by geometry, kinematics and coupling. In the limit of
small velocities and weak gravitational fields, the extra-acceleration of a
dust fluid is given by
\begin{equation}
\vec{a}_{E}=-\lambda \nabla \left[ f_{2}(R)\right]. %
\end{equation}

The thermodynamic condition for the vanishing of the extra-force is $%
\partial L_{m}/\partial g^{\mu \nu }=\left( 1/2\right) \left( \partial
L_{m}/\partial \rho \right) \rho \left( g_{\mu \nu }-u_{\mu }u_{\nu }\right)
=0$ only. If the matter Lagrangian is written as a function of the pressure,
then $\partial L_{m}/\partial \rho =\left( \partial L_{m}/\partial p\right)
\left( \partial p/\partial \rho \right) $, and for all physical systems
satisfying an equation of state (or, equivalently, for all systems with a
non-zero sound velocity), the extra-force is non-zero. Therefore, the
geometry-matter coupling is introduced in the generalized gravity models \cite
{Bertolami:2007gv} and \cite{ha08} in a consistent way. The coupling
determines all the physical properties of the system, including the
extra-force, the matter Lagrangian, and the energy-momentum tensor, respectively.

\section*{Acknowledgments}

This work is supported by an RGC grant of the government of the Hong Kong
SAR.

\section*{Appendix: Proof of the relation $\delta \rho =%
\frac{1}{2}\rho \left( g_{\mu \nu }-u_{\mu
}u_{\nu }\right) \delta g^{\mu \nu }$}

Let's consider a fluid with density $\rho $ in a system of arbitrary
coordinates $x^{\lambda }$. Let's introduce an arbitrary coordinate
transformation to a new coordinate system $a^{\mu }$, so that $x^{\alpha
}=f^{\alpha }\left( a^{\lambda }\right) $, $\alpha =0,1,2,3$, $\lambda
=0,1,2,3$. In the new coordinate system $a^{\lambda }$ the components of the
metric tensor $\eta _{\mu \nu }\left( a^{\lambda }\right) $ are given by $%
\eta _{\mu \nu }\left( a^{\lambda }\right) =g_{\alpha \beta }\left(
x^{\lambda }\right) \left( \partial f^{\alpha }/\partial a^{\mu }\right)
\left( \partial f^{\beta }/\partial a^{\nu }\right) $. We define the
four-velocities of the fluid in the two frames as $u^{\alpha }=dx^{\alpha
}/ds_{(x)}$ and $Q^{\alpha }=da^{\alpha }/ds_{(a)}$, where $%
ds_{(x)}^{2}=g_{\alpha \beta }dx^{\alpha }dx^{\beta }$ and $%
ds_{(a)}^{2}=\eta _{\alpha \beta }da^{\alpha }da^{\beta }$, respectively.
The fluid obeys the equations of mass continuity, which in the two reference
frames are given by
\begin{equation}
\frac{1}{\sqrt{-g}}\frac{\partial }{\partial x^{\alpha }}\left( \sqrt{-g}%
\rho u^{\alpha }\right) =0,\frac{1}{\sqrt{-\eta }}\frac{\partial }{\partial
x^{\alpha }}\left( \sqrt{-\eta }\rho Q^{\alpha }\right) =0.  \label{a1}
\end{equation}
In Eqs.~(\ref{a1}) we have taken into account that $\rho $ is a scalar
quantity, that is, it is invariant with respect to this coordinate
transformation. Let's now assume that the coordinate transformation $%
f^{\alpha }$ takes our current coordinates of the particles in the fluid, $%
x^{\lambda }$, to the initial coordinates of the particles $a^{\lambda }$
(the Lagrange coordinates). The corresponding Lagrange coordinate system is
a comoving coordinate system - in it the fluid appears to be at rest. In
Lagrange coordinates the space-like coordinates $a^{k}$ are constants, $%
a^{k}=a^{k}(0)$, $k=1,2,3$. The fluid velocity in the comoving coordinate
system is
\begin{equation}
Q^{\alpha }=\frac{da^{\alpha }}{ds_{(a)}}=\delta _{0}^{\alpha }\left[
g_{\alpha \beta }\left( f^{\lambda }\right) \frac{\partial f^{\alpha }}{%
\partial a^{0}}\frac{\partial f^{\beta }}{\partial a^{0}}\right] ^{-1/2}.
\end{equation}

In the Lagrangian coordinates $a^{\lambda }$ the equation of continuity
becomes
\begin{equation}
\frac{\partial }{\partial a^{0}}\left[ \sqrt{-\eta }\rho \left( g_{\alpha
\beta }\left( f^{\lambda }\right) \frac{\partial f^{\alpha }}{\partial a^{0}}%
\frac{\partial f^{\beta }}{\partial a^{0}}\right) ^{-1/2}\right] =0,
\end{equation}
and can be integrated immediately to give
\begin{equation}
\sqrt{-\eta }\rho =\sqrt{g_{\alpha \beta }\left( f^{\lambda }\right) \frac{%
\partial f^{\alpha }}{\partial a^{0}}\frac{\partial f^{\beta }}{\partial
a^{0}}}F\left( a^{1}(0),a^{2}(0),a^{3}(0)\right) ,
\end{equation}
where $F$ is an integration function, depending on the Lagrange coordinates $%
a^{k}$, $k=1,2,3,$ only. Taking into account that $\sqrt{-\eta }=\sqrt{-g}%
\left[ \partial (x)/\partial (a)\right] $, where $\partial (x)/\partial (a)$
is the absolute value of the Jacobian of the coordinate transformation $%
x^{\alpha }\rightarrow a^{\alpha }$, we obtain
\begin{equation}
\rho \sqrt{-g}\frac{\partial (x)}{\partial (a)}=\sqrt{g_{\alpha \beta
}\left( f^{\lambda }\right) \frac{\partial f^{\alpha }}{\partial a^{0}}\frac{%
\partial f^{\beta }}{\partial a^{0}}}F\left(
a^{1}(0),a^{2}(0),a^{3}(0)\right) .  \label{a2}
\end{equation}
In the $a^{\alpha }$ coordinates the velocity four-vector takes the form
\begin{equation}
u^{\alpha }=\frac{df^{\alpha }}{ds_{(a)}}=\left( \frac{\partial f^{\alpha }}{%
\partial a^{0}}\right) \left( g_{\mu \nu }\left( f^{\lambda }\right) \frac{%
\partial f^{\mu }}{\partial a^{0}}\frac{\partial f^{\nu }}{\partial a^{0}}%
\right) ^{-1/2}.
\end{equation}
Since the Jacobian $\partial (x)/\partial (a)$, as well as the function $%
F\left( a^{1}(0),a^{2}(0),a^{3}(0)\right) $, does not contain the quantities
$g_{\mu \nu }$, the variation with respect to $g_{\mu \nu }$ in Eq.~(\ref{a2}%
) gives
\begin{equation}
\frac{\partial (x)}{\partial (a)}\delta \left( \rho \sqrt{-g}\right) =\frac{1%
}{2}F\sqrt{g_{\mu \nu }\left( f^{\lambda }\right) \frac{\partial f^{\mu }}{%
\partial a^{0}}\frac{\partial f^{\nu }}{\partial a^{0}}}u^{\mu }u^{\nu
}\delta g_{\mu \nu }.
\end{equation}

But
\begin{equation}
F\sqrt{g_{\mu \nu }\left( f^{\lambda }\right) \frac{\partial f^{\mu }}{%
\partial a^{0}}\frac{\partial f^{\nu }}{\partial a^{0}}}=\rho \sqrt{-g}\frac{%
\partial (x)}{\partial (a)},
\end{equation}
and therefore
\begin{equation}
\delta \left( \rho \sqrt{-g}\right) =\frac{1}{2}\sqrt{-g}\left( \rho u^{\mu
}u^{\nu }\right) \delta g_{\mu \nu }=-\frac{1}{2}\sqrt{-g}\left( \rho u_{\mu
}u_{\nu }\right) \delta g^{\mu \nu }.
\end{equation}

Combining this equation with the known relation $\delta \sqrt{-g}=-\left(
1/2\right) \sqrt{-g}g_{\mu \nu }\delta g^{\mu \nu }$ gives
\begin{equation}
\delta \rho =\frac{1}{2}\rho \left( g_{\mu \nu }-u_{\mu }u_{\nu }\right)
\delta g^{\mu \nu }.
\end{equation}

\end{document}